\documentclass[fleqn,10pt]{wlscirep}
\usepackage{ragged2e}

\usepackage{multirow}

\title{Predictive Functional Connectivity of Real-World Systems}

\author[1,$\dag$]{Anida Sarajli\'c}
\author[1,$\dag$]{No\"el Malod-Dognin}
\author[2]{\"Omer Nebil Yavero\u{g}lu}
\author[1,*]{Nata\v{s}a Pr\v{z}ulj}
\affil[1]{Department of Computing, Imperial College London, SW7 2AZ London, United Kingdom}
\affil[2]{Google UK, London, United Kingdom}

\affil[$\dag$]{These authors contributed equally}
\affil[*]{natasha@imperial.ac.uk}

\begin{abstract}
We are flooded with large-scale dynamic networked data. Analyses requiring exact comparisons between networks are computationally intractable, so new methodologies are sought.  We extend the graphlet-based statistics to directed networks and demonstrate that they are superior to other measures. We predict a country's gross domestic product (GDP) solely from its wiring patterns in the world trade network (WTN) that could inform policy makers on benefits of trade agreements.  Surprisingly, we find that it is not enough for a country to be in a densely connected core in the WTN to have a high GDP, as was previously believed. In addition to being in the core, a country must also trade with peripheral countries, while only being in the core and not trading with peripheral economies makes a country prone to debt.  Furthermore, by tracking the dynamics of a country's positioning in the WTN over years, we predict success or failure of an emerging economy. We validate our methodology on metabolic networks, yielding insights into preservation of enzyme function from the network wiring patterns rather than sequence data.  Overall, our methodology enables advanced analyses of directed networked data from any area of science, allowing domain-specific interpretation of a directed network's topology.\\

Supplementary material is available at \url{http://bio-nets.doc.ic.ac.uk/Directed/Supplement.pdf}
\end{abstract}
\begin{document}
\flushbottom
\maketitle

%
\thispagestyle{empty}

\section*{Introduction}
Deciphering the wiring patterns (also called topology) of large-scale, dynamic, directed, networked data coming from all domains of sciences is fundamental for understanding and predicting their functioning, emergent properties and controllability{\cite{Barabasi-Nat-473,Galbiati13}}.
Topological analyses of networks are universal and provide insights in all areas of science that use network theory, including social science{\cite{scott2012social}}, politics{\cite{ward2011network}}, biology{\cite{junker2011analysis}} and medicine{\cite{barabasi2011network}}.

In economics, yearly trade relationships between countries are captured in the world trade network (WTN), in which nodes represent countries and in which there is a directed edge from country $u$ to country $v$ 
if $u$ exported some commodities (i.e., marketable items) to $v$ in the considered year.
In the literature, topological analyses of these directed networks focused on characterizing their organization.
Garlaschelli and Loffredo {\cite{garlaschelli2005}} were the first ones to hypothesise the modular organization of WTN, due to its similarities with scale-free networks{{\cite{barabasi99}}. They  showed that similar to scale free topology, WTN is characterized by a sharp power-law distribution of directed edges (e.g., many countries are involved in a few trades but only few countries are involved in large numbers of trades) and that trade relationships are disassortative (e.g., countries with many trade partners are connected to countries with few partners and vice-versa). 
This modular organization of WTN was refined by Kali and Reyes {\cite{kali2007}}, who characterized the core-periphery organization {\cite{borgatti2000models}} of trades using in-degree centrality: in the WTN, some countries are at the densely connected core (the central nodes in the WTN), forming rich-clubs of trading countries, while others (the non-central ones) are at the sparsely connected periphery.
The core-periphery organization was later observed by using various statistics such as in-degree and betweenness centralities {\cite{debenedictis2011}}, random walkers and $k$-shell decomposition {\cite{della2013profiling}}. It was also shown to be affected by trade globalization and regionalization {\cite{debenedictis2011}}.
These studies also highlighted the link between the positioning of a country in the WTN and its economy. For example, Garlaschelli and Loffredo {\cite{garlaschelli2005}} proposed to predict the gross domestic product (GDP) of a country by a derivative of its number of trading partners (i.e., node degree).

In systems biology, metabolism is the set of all life-sustaining chemical reactions in a cell. These reactions, in which enzymes catalyse the transformation of substrates (input metabolites) into products (output metabolites), can be represented by metabolite-centred and by enzyme-centred metabolic networks. In metabolite-centred metabolic networks, two metabolites (nodes), $u$ and $v$, are connected by a directed edge from $u$ to $v$ if there is an enzyme that catalyse a reaction having $u$ as one of its substrates and $v$ as one of its products. In enzyme-centred metabolic networks, two enzymes (nodes), $u$ and $v$, are connected by a directed edge from $u$ to $v$ if some of the products of $u$ are substrates of $v$. The topology of metabolic networks has been the focus of many studies (e.g., the review of Lacroix {\em et al.}{\cite{lacroix2008introduction}}). In these studies, a popular network statistic is {\em network motifs}{\cite{milo2002network}}, which are defined as small partial sub-graphs that are over-represented in a network with respect to a given null model (partial sub-graph means that when selecting a sub-set of nodes from the large network, we can select any edges between them).

Similar to WTN, many studies focused on characterizing the organization of metabolic networks.
Zhu and Qin {\cite{zhu2005structural}} observed that metabolic networks have scale free topology with highly modular organization, coming from their power law distributions of in- and out-degrees.
Recently, Shellman {\em et al.}{\cite{shellman2013network}} showed that this modular organization is related to the cellular organization: while the distributions of 3-node motifs in metabolic networks are similar across species, they are unique across organelles (i.e., different cellular components).
Furthermore, the topological similarities between the metabolic networks of different species are shown to approximately reconstruct phylogenetic classification of species {\cite{heymans2003deriving,zhang2006phylophenetic}}, which suggests a link between the divergence times between species and the dissimilarities between their metabolic networks. However, this observation must be toned down by the fact that metabolic networks are reconstructed from homology{\cite{francke2005reconstructing,feist2009reconstruction}}. 
Recently, Pearcy {\em et al.}{\cite{pearcy2015network}} related the similarities between the metabolic networks of 383 bacterial species, as measured by the similarity of their motif spectra, with their phenotypic variability (e.g., aquatic or terrestrial species, aerobic or anaerobic environment), which suggests that adaptation to environment during evolution may have shaped the topology of metabolic networks.
Finally, motifs and their spectra were used to relate the positioning of enzymes and metabolites in metabolic networks and their biological functions. 
Shellman {\em et al.}{\cite{shellman2013network}} showed that some motifs are characteristic to specific metabolic functions of enzymes, and going further, Ganter {\em et al.}{\cite{ganter2014predicting}} proposed a hidden markov model{\cite{ghahramani2001introduction}} based framework for predicting metabolic functions from motif spectra.

Because exact comparison between complex networks has long been known to be computationally intractable {\cite{cook71}}, the topological analyses of complex networks use simple heuristics, commonly called network \emph{statistics}, such as the in- and out-degree distributions, to approximately say whether the structures of networks are similar {\cite{newman2009networks}}.
Recently, the concept of \emph{node roles}, which associate nodes with well defined topological features, has been proposed for analysing networked data.
For example, in control theory, the set of driver nodes that can control and move the networks into specific states, has been identified and shown to be of low degree {\cite{Barabasi-Nat-473}}.
In the same vein, Yan {\em et al.}{\cite{yan2015spec}} classify nodes in network as indispensable, neutral or dispensable, if their removal from the networks results in increasing, similar, or decreasing number of driver nodes in the resulting networks, respectively. In directed protein-protein interaction networks, indispensable proteins are shown to be the primary targets of disease-causing mutations, and this observation was used to predict novel cancer-driver genes.
However, the limited number of controllability-based roles limits their interpretability.
As seen above, nodes in metabolic networks can be characterized by their motif spectra{{\cite{milo2002network}}}. These motif-based roles are used to relate the positioning of enzymes in metabolic networks with their biological functions{\cite{shellman2013network,ganter2014predicting}}. However, motifs and their spectra have limited usability{\cite{artzy2004comment}}, as they are dependant on the choice of a null model, which is generally unknown for real-world data.

For analysing undirected networks, Yavero\u{g}lu {\em et al.} {\cite{yaveroglu14}} proposed a framework in which node roles are modelled by graphlets{\cite{przulj04}}.
Graphlets are defined as small induced subgraphs of a large network that appear at any frequency; an induced sub-graph means that once you pick the nodes in the large network, you must pick all the edges between them to form the sub-graph. Within graphlets, \emph{symmetry groups} of nodes called \emph{automorphism orbits} are used to characterize different topological positions that a node can participate in. Orbits are used to generalize the notion of node degree: the {\em graphlet degrees} of a node are the numbers of times a node is found at each orbit position {\cite{przulj07}}.
Graphlets with up-to five nodes and their degrees characterize 73 different node roles, and their interpretability is independent of a null model.
In the node role framework, the dependencies between node roles (i.e., the correlations between the counts of graphlet orbits over all nodes in a network) are encoded in a symmetric matrix called the graphlet correlation matrix (GCM), which is shown to finely describe the topology of undirected networks{\cite{yaveroglu14}}. GCMs have also been compared within the network statistics called graphlet correlation distance (GCD), which has been shown to be the most accurate network statistic for classifying undirected networks. Finally, canonical correlation analysis (CCA){\cite{hair2006multivariate}} allows translating node roles in network into domain specific languages. In economics, CCA is used to relate the graphlet-based roles of countries in the undirected world trade network with their economic attributes{\cite{yaveroglu14}}.
In systems biology, this framework is used to uncover biological functions that are performed through similar patterns of protein-interactions across yeast and human (topologically orthologous biological functions){\cite{davis2015topology}}.

However, graphlets, and by extension, the whole node role framework, are only defined for undirected networks. Some real-world data are inherently directed and cannot be represented by undirected networks without information loss. For example, a trade between two countries is directed, from the exporting country towards the importing country, and these two situations are not equivalent; the trade represents an income for the exporting country, while it represents an expense for the importing country. When world trades are modelled as undirected networks, directionality is lost and the two situations are treated equally.

%
%
To better analyse directed data and to avoid information loss from modelling directed data with undirected networks, we extend the node role framework to directed networks.
First, we define directed graphlets, and extend existing graphlet-based statistics to directed networks. Among them, we show that the directed graphlet correlation distance is superior for clustering directed networks.
Second, we use our directed graphlets to extend the node role framework to directed networks, and demonstrate its superior descriptive and predictive power on real-world data in two domains, economics and biology.
%
%

\section*{New methodology: node role descriptors for directed networks}
To adapt the node role framework to directed networks, we first generalize graphlets to directed ones as follows.
We define directed graphlets as small induced sub-graphs of a larger directed network, without anti-parallel directed edges (i.e., if a graphlets contains a directed edge from node $u$ to node $v$, it cannot contain the opposite directed edge from node $v$ to node $u$).
The 40 two-to-four node directed graphlets are presented in Fig. \ref{fig:graphlets} (panel a) and are denoted from $G_0$ to $G_{39}$; their 129 automorphism orbits are denoted from 0 to 128.
Real-world directed networks may contain anti-parallel directed edges, which we take into account in our counting strategy by inducing directed graphlets for each anti-parallel directed edge separately (Fig. \ref{fig:graphlets}, panel c).
Analogous to the graphlet degree vector {\cite{milenkovic08}}, a node in a directed network is described by its {\em Directed Graphlet Degree Vector} (DGDV), which is a 129-dimensional vector encoding the two- to four-node graphlet degrees of the node in the networks; e.g., the $i^{th}$ coordinate of the DGDV of node $n$, denoted by $DGDV_n[i]$ is the number of times a directed graphlet touches node $n$ at orbit $i$.

\subsection*{Directed graphlet-based statistics}
Directed graphlets are like Lego pieces that assemble with each other to build large networks; any directed network can be constructed by combining different directed graphlets at different directed-orbits (an example is given in Fig \ref{fig:graphlets}, panel b). We exploit this observation to summarize the complex structures of networks and to compare them, by generalizing three graphlet-based network statistics to directed networks: the Relative Graphlet Frequency Distribution Distance{\cite{przulj04}}, the Graphlet Degree Distribution Agreement{\cite{przulj07}} and the Graphlet Correlation Distance{\cite{yaveroglu14}}.
{\em Directed Relative Graphlet Frequency Distribution Distance} (DRGF) compares two networks according to their relative distributions of directed graphlet frequencies. Given the directed graphlet frequency vectors of two networks, DRGFD first normalizes these vectors based on the total number of directed graphlets that appear in the networks. Then, it computes the sum of absolute differences between the normalized directed graphlet frequencies. The resulting score measures the topological difference between the two networks (see Supplementary material).
{\em Directed Graphlet Degree Distribution Agreement} (DGDDA) compares two networks according to their distributions of directed graphlet degrees (directed graphlet degree is number of times the node touches a specific directed orbit). Given two networks, DGDDA method compares pairs of directed orbit degree distributions separately and quantifies the overall topological similarity between the two networks as an average over all comparisons (see Supplementary material).
In {\em Directed Graphlet Correlation Distance} (DGCD), dependencies between directed graphlet orbits inside a network are summarized into an $n\times n$ {\em Directed Graphlet Correlation Matrix} (DGCM), where $n$ is the number of considered directed graphlet orbits. Each entry of this matrix quantifies the dependency between two graphlet orbits in the network, which we measure by Spearman's correlation{\cite{spearman1904}} between the corresponding directed graphlet degrees over all nodes in the network. In addition to in-depth examination of network topology that can be qualitatively interpreted (see section Results), we exploit the differences in DGCMs to compare directed networks. DGCD measures the distance between networks as the Euclidean distance between their DGCMs (see Supplementary material). Its superiority over other directed network distances is demonstrated in section Evaluation. Note that we consider two versions of DGCD: DGCD-129, which uses the 129 two- to four-node directed graphlet orbits, and DGCD-13, which uses only the 13 two- to three-node directed graphlet orbits.

Directed graphlets can also be used to measure the similarity between the positioning of nodes in directed networks; the {\em Directed Graphlet Degree Vector Similarity} (DGDVS) measures the similarity between the positioning of two nodes by the similarity of their directed graphlet degree vectors (see details in Supplementary material).

\subsection*{Directed node role framework}
To bridge the gap between node roles (the wiring patterns around nodes that are captured by their directed graphlet degree) in a network and their (real-valued) domain specific annotations, we adapt the canonical correlation analysis (CCA) methodologies from Yavero\u{g}lu {\em et al.}{\cite{yaveroglu14}} and from Davis {\em et al.}{\cite{davis2015topology}} to uncover the linear relationships between them.

In our CCA framework, the node roles are encoded in a node role matrix, $R^{t}$, and the domain specific annotations in an attribute matrix, $R^{f}$.
For both matrices, each row (called a {\em variable vector}) represents a specific node (for both matrices, the same row index represents the node).
In the node role matrix, values in a variable vector (a row) are the 129 directed graphlet degrees of the node in the network, while in the attribute matrix, they correspond to the annotations of the node.

Given $n$ pairs of variable vectors from $R^{t}\times R^{f}$ for $n$ nodes as an input, the CCA outputs two weight
vectors, called a {\em canonical variate}, so that the weighted sum of $R^{t}$ is maximally correlated with the weighted sum of $R^{f}$. The correlation between the two weighted sums is called their {\em canonical correlation}. After finding the first set of weights, CCA iterates $min\left \{ t,f \right \}$ times to find more weight vectors, such that the resulting canonical variates are not correlated with any of the previous canonical variates.
The weight matrices $W_{1}$ and $W_{2}$, for $R^{t}$ and $R^{f}$, respectively, are constructed by combining all of the identified weight vectors.
We refer the interested reader to Weenink{\cite{weenink2003canonical}} for the mathematical aspects of CCA.

For a given canonical variate, the contribution of a variable (e.g., of a given directed graphlet degree) is measured by its {\em loading}, which is the Pearson's correlation{\cite{pearson1895}} between the values of the variable and of the linear combination of variables that it participates in.
For each canonical variate, the variables having the largest (positive or negative) loadings are the most related to each others, which allows uncovering the strongest relationships between node roles and annotations.

Finally, \emph{the association matrix}, which is constructed as $ W_{1} \times  S \times W_{+}^{2}$, where $S$ is the diagonal matrix of canonical correlations that weights the variates according to their correlation strength and where $W_{+}^{2}$ is the Moore-Penrose pseudoinverse\cite{albert1972regression} of $W_{2}$, is a multidimensional regression that allows predicting the domain specific annotations of a node from its directed graphlet degrees in the network.

\section*{Evaluation of the methodology}
We compare the performance of our novel directed graphlet-based network distances (DGCD-129, DGCD-13, DRGF and DGDDA) used for clustering networks coming from six directed network models (ER, SFBA-Sink, SFBA-Source, GEO, GEO-GD and SF-GD), which we define as follows.
\begin{itemize}
\item {\em Directed Erd\"{o}s-R\`{e}nyi random model, {\bf \em ER}}, represents uniformly distributed random interactions {\cite{erdos59}}. A directed ER network is generated by fixing the number of nodes in the network, and then by randomly adding directed edges between uniformly chosen pairs of nodes.
\item {\em Scale free Barab\`{a}si-Albert random model,} also called preferential attachment, generate networks based on the ``rich-gets-richer'' principle{\cite{barabasi99}}.
We use the directed preferential attachment model{\cite{bollobas03}} to generate two distinct scale free topologies. In the first one, {\bf \em SFBA-Sink}, we favour adding directed edges towards nodes having already large numbers of incoming edges, creating densely connected sink nodes. In the second one, {\bf \em SFBA-Source}, we favour adding directed edges outgoing from nodes having already large numbers of outgoing edges, creating densely connected source nodes.
\item {\em Directed scale-free random model with gene duplication and divergence, {\bf \em SF-GD}}, is a scale-free model which mimics the gene duplication and the gene divergence processes from biology~{\cite{vazquez02}}. Starting from a small seed network, a node in the network is selected at random and a new node is created together with the connections to/from nodes that the ``parent'' node has (duplication step).
A directed edge, whose directionality is randomly chosen, is added between the new node and his parent with probability $p$.
Then, each directed edge that the new node ``inherited'' from its parent node is deleted with probability $q$ (divergence step).
This procedure is repeated until the desired number of nodes is achieved. Probabilities $p$ and $q$ are tuned to achieve the desired number of directed edges.
\item {\em Directed geometric random model, {\bf \em GEO}}, represents the proximity relationship between uniformly randomly distributed points in a $d$-dimensional metric space~{\cite{penrose03}}.
We generate GEO networks by putting points in 3-dimensional Euclidean unit cube uniformly at random that correspond to nodes of the networks
and two nodes are connected by a directed edge, whose directionality is randomly chosen, if the Euclidean distance between the corresponding points in space is smaller than a specified distance threshold, $r$, which is chosen to achieve the desired number of directed edges.
\item {\em Directed geometric random model with gene duplication, {\bf \em GEO-GD}}, is a geometric model that mimics the gene duplication and divergence processes in biology~{\cite{przulj10}}. Starting from a small number of nodes randomly distributed in a metric space, a new node is introduced as a copy of a randomly selected node in the network (duplication step); the new node is moved randomly from its \emph{parent} node in the metric space (divergence step), at a random distance smaller or equal to $2 \times r$ (where $r$ is the distance threshold used to generate GEO networks). This process is repeated until the desired number of nodes is achieved, and then directed edges between nodes are created following the rules of the directed geometric network model.
\end{itemize}

We extend the methodology of Yavero\u{g}lu {\em et al} {\cite{yaverouglu2015proper}}.
We generate 10 networks for each model and for each of the following three numbers of nodes and two directed edge densities that mimic the sizes and densities of real-world networks: 500, 1000 and 2000 nodes, and 0.5\% and 1\% directed edge densities. Hence, the total number of synthetic networks that we consider is $10\times 6\times 3\times 2 = 360$.
We formally evaluate the performance of our directed graphlet-based network distances for grouping networks from these models together and compare them to the performances of other commonly used directed network distance measures: the in- and out-degree distribution distances (note that we only present results of in-degree distribution distance, as it produces almost identical results as its out-degree counterpart) and directed spectral distance {\cite{wilson08}}. We assess how well a distance measure groups networks of the same type by using the standard Precision-Recall and ROC curves {\cite{fawcett06}}: for small increments of parameter $\epsilon \geq 0$, if the distance between two networks is smaller than $\epsilon$, then the pair of networks is declared to be similar (belong to the same cluster). For each $\epsilon$, four values are computed:
the {\em true positives} (TP), which are the numbers of correctly clustered pairs (i.e., that group together networks from the same model),
the {\em true negatives} (TN), which are the numbers of correctly non-clustered pairs (i.e., that do not group together networks from different models),
the {\em false positives} (FP), which are the numbers of incorrectly clustered pairs (i.e., that group together networks networks from different models),
the {\em false negatives} (FN), which are the numbers of incorrectly non-clustered pairs (i.e., that do not group together networks from the same model).
In {\em Precision-Recall curves}, for each $\epsilon$, the precision ($PREC=\frac{TP}{TP+FP}$) is plotted against the recall, also-called true-positive rate, ($REC=\frac{TP}{TP+FN}$) and the quality of the grouping by a given distance measure is measured with the area under the Precision-Recall curve (AUPR), which is the average precision of the distance measure.
In {\em ROC curves}, for each $\epsilon$, the true-positive rate is plotted against the false-positive rate ($FPR=\frac{FP}{FP+TN}$) and the quality of the grouping by a given distance measure is measured with the area under the ROC curve (AUC), which is the probability that a randomly chosen pair of networks coming from the same model will have a distance smaller than a randomly chosen pair of networks coming from different models.

AUPRs and AUCs show that DGCD-13 is the most accurate among all tested measures (Fig. \ref{fig:clustering}, panel a), which is also illustrated by its superior Precision-Recall curve (Fig. \ref{fig:clustering}, panel b).
The superiority of DGCD-13 over DGCD-129 may come from the large number of dependencies between three-nodes and four-nodes directed graphlet degrees (listed in the Supplementary material), which may blur the (dis)similarities between the DGCMs used by DGCD-129.
Since the closest objects are the first to start forming clusters, we are interested in distance measures that optimize the number of correctly clustered pairs of networks that are at the shortest distances and thus, that are retrieved first by the distance measure {\cite{yu2006}}.
Both DGCD-13 and DGCD-129 show superiority in early retrieval over all measures (beginnings of the curves in Fig. \ref{fig:clustering}, panel b).
Furthermore, we confirm that such results could not have been achieved by undirected network measures by comparing the performances of directed distance measures to the ones of their undirected counterparts (when applied on the same networks but that are made undirected by removing the directionality of the edges). As expected, undirected network distances perform poorly, as evidenced by smaller AUCs and AUPRs (Fig. \ref{fig:clustering}, panel a) and 
Precision-Recall curves (Fig. \ref{fig:clustering}, panel c).

Since real networks are incomplete and noisy, we evaluate the clustering quality of the above distance measures in the presence of noise. To simulate noise, we randomly rewire up to 90\% of the directed edges of model networks, in increments of 10\%. To account for the variability of randomizations, for each percentage of noise we repeat the randomisation process 30 times and report minimum, average and maximum AUPR for clustering networks from the same model. We follow the same procedure to simulate missing data by randomly removing up to 90\% of the directed edges from model networks, in increments of 10\%.
These tests demonstrate robustness to noise and missing data and the superiority of DGCD over other measures on this sample of different network topologies, sizes and densities (Fig. \ref{fig:clustering}, panel d-e).

We further validate our methodology by assessing its ability to correctly group real-world metabolic networks.
We reconstruct the directed metabolic networks of all Eukaryotes (299 species) using the metabolic reaction data from KEGG database {\cite{kanehisa96}} (collected in December 2014). For each species, we model its metabolic reactions as an enzyme-enzyme network in which two enzyme-coding genes (nodes) are connected by a directed edge if the first enzyme catalyses a reaction whose product is a substrate for a reaction catalysed by the second enzyme.
We obtain the taxonomic classification of species from KEGG database, which classifies Eukaryotic species according to (from the most generic to the most specific) their Kingdom, Sub-phylum and Class.
We use the six directed distance measures to compare our 299 directed metabolic networks and use Precision-Recall and ROC curve analyses to measure their agreement with their taxonomic classification. Since the metabolic reactions encoded in metabolic networks are mostly reconstructed from sequence homology relationships, we expect that the similarity between the topologies of our directed metabolic networks will relate to the evolutionary relationships between the corresponding species, validating our methodology.
Indeed, as detailed in supplementary Fig. 3, DGCD-129 and DGCD-13 achieve the highest agreements with the taxonomic classification of species.
For example, DGCD-13 of directed metabolic networks agree with their Class classification with AUC of 0.945, validating the performance of the methodology.

\section*{Application to real-world networks}

\subsection*{Economics: world trade network}
Using trade data from the United Nations Commodity Trade Statistics (UN Comtrade) database (collected in August 2014, from \url{http://comtrade.un.org/}), we generate 52 WTNs, one for each year between 1962 and 2013. To filter-out insignificant trades, we only consider the largest trades accounting for 90\% of the total trade amount of a given year. We obtain the economic indicators of country wealth from PENN World Table (PENN, version 7.1, downloaded in July 2014 from \url{https://pwt.sas.upenn.edu/}) and International Monetary Fund (IMF) World Economic Outlook Database (WEO, downloaded in July 2014 from \url{http://www.imf.org/}).
For simplicity, we focus on the following nine economic indicators of countries: population size, employed population size, gross domestic product (GDP), GDP per capita (i.e., GDP of a country divided by its population size), debt, debt per capita (i.e., debt of a country divided by its population size), import expenses, export incomes  and capital stock. Because of the limited availability of economic indicators, we do our analyses linking a country's wiring patterns in the WTN to its economic indicators on the subset of 32 WTNs from 1980 to 2011.

\paragraph{Uncovering the directed core-broker-periphery structure of WTN.}
We use the descriptive power of the directed graphlet correlation matrices (DGCMs) to analyse the structure of directed WTNs.
Yavero\u{g}lu {\em et al.}\cite{yaveroglu14} highlighted the importance of broker positions in the undirected WTN, which mediate the trade between otherwise disconnected peripheral countries and the core of densely connected countries, leading to the core-broker-periphery model of undirected WTN.

Each of graphlets $G_{18}$ to $G_{25}$ (Fig. \ref{fig:graphlets}, panel a) contains all three positions; the core is represented by the triangle's nodes (e.g., orbits 49, 50 and 51 on graphlet $G_{18}$), the periphery is represented by the node hanging from the triangle (e.g., orbit 52 on graphlet $G_{18}$), and the broker by the node from the core that is connected to the peripheral node (e.g., orbit 51 on graphlet $G_{18}$).
We use these eight graphlets to assess if our directed WTN also possess the core-broker-periphery organization. To this aim, we focus on the DGCMs that are obtained when considering three sets of orbits: the eight peripheral orbits (orbits 52, 56, 60, 64, 68, 72, 76 and 80); the eight broker orbits (orbits 51, 55, 59, 63, 67, 71, 75 and 79); and the 16 core orbits that are not brokers (orbits 49, 50, 53, 54, 57, 58, 61, 62, 65, 66, 69, 70, 73, 74, 77 and 78). These orbits are illustrated in Fig. \ref{fig:dwtn_struct}, panel a.

In our WTNs of each year, as presented in Fig. \ref{fig:dwtn_struct} (panels b, c and d), the broker orbits strongly correlate with the core orbits, showing that core trading countries are very likely to be involved in trade brokerage. On the opposite, the peripheral orbits are not correlated with core orbits and slightly anti-correlated with brokerage orbits, showing that in the WTN there is a clear distinction between core/broker and peripheral countries.
Thus, similar to the undirected case, our directed WTNs are characterized by a core-broker-periphery structure.
Moreover, we observe that peripheral countries form two subsets: the peripheral countries that import from the broker countries (orbits 52, 60, 68 and 76) and peripheral countries that export to the broker countries (orbits 56, 64, 72 and 80). The strong correlation within these two sets and the low correlation between them show that the peripheral countries tend to specialize as either peripheral importer or as peripheral exporter, but not both.
This observation is only made possible by our new directed graphlet-based analysis and has eluded us thus far in undirected and directed studies.
Over time, the correlations between the peripheral roles and the core/broker roles slightly increases, showing that peripheral countries are getting more integrated into the WTNs, which is likely to be an effect of globalization.
However, while the correlation within the two sets of peripheral orbits get stronger over time, the correlations between them get weaker, which shows that peripheral countries became more specialised towards either import or export.
Finally, the directed core-broker-periphery structure of WTNs does not appear in random networks (e.g., see the DGCM of a directed Erd\"os-R\`enyi random network in Fig. \ref{fig:dwtn_struct} panel e).

\paragraph{Predictive power of directed wiring patterns.}
We use our new CCA framework to compute the association matrix transforming the directed graphlet degrees of a country in the WTN into predicted economic indicators, using all countries and all years from 1980 to 2011 simultaneously.
We construct a single node role matrix, $R^{t}$, and a single economic attribute matrix, $R^{f}$, which encompass all countries and all years, as follows.
For both matrices, each row represents a country for a given year; e.g., their is one row for USA in 1980, one row for USA in 2000, one for UK in 2010... 
In the node role attribute matrix, $R^{t}$, values in a variable vector are the 129 directed graphlet degrees of the country in the corresponding world trade network, while in the economic attribute matrix, $R^{f}$, they correspond to the nine economic attributes of the country in that year.

We apply our CCA framework on these two matrices, and first assess the ability of the produced association matrix to predict the economic attributes of the countries from their node roles in the WTNs, which we measure by the Pearson's correlations between predicted and observed economic indicator values for each year separately.
%
As presented in Figure \ref{fig:dwtn} (panel a), capital stock, import expenses, export incomes, and GDP are all consistently well-predicted from trade patterns over the years.
In contrast, debt and population-based economic indicators (population size, employment, GDP per capita, debt and debt per capita) are not consistently predicted, showing that these indicators do not depend only on trade patterns, but also on other geopolitical factors; for example, debt is worst predicted from WTN topology in 1991, which corresponds to the Gulf War in the Middle East, the beginning of the Yugoslav Wars in eastern Europe and the Japanese asset price bubble collapse (the starting point of the ``Lost Decade'' of stagnation of Japanese economy) in East Asia.

In the second step, we measure the quality of the predicted economic indicators by the Pearson's correlations between predicted and observed economic indicator values over all years simultaneously, and compare the quality of the obtained predictions to the ones that are obtained when WTNs are made undirected (by removing directionality of the edges) and when node roles are captured by undirected graphlet degrees.
As presented in Figure \ref{fig:dwtn} (panel b), using directed node roles always results in better predictions of economic indicators, justifying the introduction of directed node roles methodology.

Among the many economic descriptors, GDP is of foremost importance, as it directly characterizes economic crises (e.g., the definition of global downturns {\cite{freund2009}}). Out of all directed orbits, orbit 71 correlates the most strongly with GDP, having Pearson's correlation of 0.88. Using all directed orbits allows predicting GDP with Pearson's correlation of 0.97! These results, providing a strong evidence of the link between the GDP and a country's wiring in the world trade network, cannot be obtained by simpler, non-graphlet-based, measures of node wiring such as in- and out-degrees, or in- and out-edge closeness centralities that were previously used for analysing directed world trade networks {\cite{debenedictis2011}}. In- and out-degrees correlate with GDP with Pearson's correlations of 0.71 and 0.72 respectively, and in- and out-edge closeness centralities with Pearson's correlations of 0.39 and 0.40 respectively.

These demonstrate that our directed graphlet-based framework finds more refined topological features than previously used methods{\cite{garlaschelli2005,debenedictis2011,yaveroglu14}}, resulting in the best predictions of the economic attributes, such as GDP of a country, from its trade wiring patterns in the WTN. Thus, our framework enables prediction of how the changes in the trade policies of a country may affect its economy.
However, despite the strong correlation between our predicted GDPs and the observed GDPs of the countries, our predictions are still not accurate enough for predicting the exact GDP values (e.g., see Supplementary Fig. 2). To further improve the quality of our predictions, additional non-trade related knowledge may be needed (e.g., external debt relations, bank exchanges or geopolitical situations). Such knowledge can be incorporated directly into our CCA framework by extending the node role matrix, $R^{t}$, with additional features, or by using data integration techniques, which are used in systems biology for fusing molecular data produced by various omics studies{\cite{gligorijevic2015integrative}}.

\paragraph{Economic interpretations of directed wiring patterns.}
We use the canonical variates (i.e., the linear combinations of graphlet degrees and their corresponding linear combinations of economic indicators) to bridge the gap between the trade patterns of the countries in the WTN and their economic attributes.

The first canonical variate, presented in Fig. \ref{fig:dwtn} (panel c), is statistically significant, with canonical correlation of 0.987 and p-value $\approx$ 0.
It highlights eight positions in the WTN enabling a country to be a middle-man in trade between non-trading countries (a broker position): orbits 55, 63, 71 and 79 correspond to countries mediating the trade between peripheral  countries exporting to trade-linked countries, and orbits 51, 59, 67 and 75 correspond to countries mediating the trade between peripheral countries importing from trade-linked countries.
In all these cases, the countries that are frequently seen in these broker positions in WTN have high GDPs and low debts by making profits from the transactions or added values by importing cheap raw materials and exporting expensive finished products, as highlighted by high export incomes and low import expenses. Also, there is a weak anti-correlation for the countries that are in peripheral positions (orbits 72, 80, 64, 56, 68, 76, 60 and 52; we only display orbits 52 and 60 in Fig. \ref{fig:dwtn} because of space limitations), indicating that these peripheral countries tend to have low GDP, low export incomes, large debt and large import expenses.

The second canonical variate, presented in Fig. \ref{fig:dwtn} (panel d), which is also statistically significant with canonical correlation of 0.95 and p-value$\approx$ 0, sheds light onto an unexpected aspect of the core-broker-periphery organisation of the WTN: countries that are in the core of the world trade but that do not trade with peripheral countries (that are not brokers, as characterized by large counts of orbits 53, 61, 77 and 69, and with small counts orbits 30 and 37) tend to have debt. This observation has completely eluded us in previous directed and undirected studies of WTN, in which being a core trading country has always been considered as a sign of economic wealth.
Our analysis suggest that in order to improve its economic wealth, a country should not only try to increase its trade relationships with the core-trading countries, but should also maximize its brokerage by trading with peripheral countries.

\paragraph{\bf Tracking world trade dynamics.}
We demonstrated  that the economy of a country is strongly related to its positioning in the WTN, highlighting favourable (broker) and dis-favourable (peripheral and core non-broker) trade positions. 
To quantify the strength of the brokerage position of a country in the WTN of each year, we extend the approach of Yavero\u{g}lu {\em et al.} {\cite{yaveroglu14}} and define the \emph{brokerage score} of the country in a particular year as the weighted linear combination of broker graphlet degrees (i.e., orbits 51, 55, 59, 63, 67, 71, 75 and 79) using the coefficients obtained from the first canonical variate.  Similarly, we quantify how \emph{peripheral} a country is in the WTN of a particular year (by using orbits 52, 56, 60, 64, 68, 72, 76 and 80). The formulas for these two scores are presented in the Supplementary material. These brokerage and peripheral scores enable us to track the changes in the position of a country in the WTN over years. We analyse if the changes in brokerage and peripheral scores of a country over years coincide with economic crises and other events impacting the economy of the country.

Indeed, we reconfirm previous observations obtained by undirected graphlet based analysis of WTN {\cite{yaveroglu14}}.
The first example is the loss of colonies of Great Britain (GBR), which was rivalling the USA as the world's leading broker of trades before decolonisation, that pushed it away from the world stage. The decline of trade brokerage position of Great Britain temporarily stabilized in 1973 when the Conservative Prime Minister, Edward Heath, led it into the European Union (EU).
However, the downward trend induced by the dissolution of the colonial superpower has continued~{\cite{kindelberg80}}.
In contrast, the reunification of Germany (DEU) in 1991 transformed it to being the central economy of Europe~{\cite{mundell2000}}.
Our scores also enable us to track the evolution of the emerging economies (supplementary Fig. 1, panel b and c).
The peripheral score of China drops after integrating Hong-Kong in 1984, which was the hub of trade between China and the rest of the world. China stopped being a peripheral trading country in 1995, after the adoption of the current foreign trade law in 1994, favouring international trade. Starting in 1995, China progressively became a broker in trade, and surpassed America in 2009, possibly benefiting from the fall of America's trade due to the global financial crisis.
To refine the above presented brokerage score, we exploit the observation that in the WTN, peripheral countries are specialized towards export or import, and define import-specific and export-specific variants of the brokerage score according to the directionality of the trade between the broker role and the peripheral role (the formulas are presented in the Supplementary material). Import- and export-specific variants of brokerage scores show that China's dominance as a broker in trade in the WTN is strongly driven by its export power, which is expected and hence validates our methodology (Supplementary Figure 1, panels c and d).

In addition, the success of China as an emerging economy can be placed in parallel with other potential emerging economies, of Brazil (BRA), Russia (RUS), India (IND), Turkey (TUR) and Indonesia (IDN). All these countries have increased their brokerage scores in the last decade, albeit not as much as China. However, they remain largely peripheral in the WTN (Fig. \ref{fig:dwtn}, panel e and f). Moreover, import- and export-specific brokerage scores (Supplementary Fig. 1, panel e and f) show that while these potentially emerging economies have increased their export-brokerage scores (with Russia and India equalling or surpassing Great Britain), their import-brokerage scores remain far behind, which may prevent them from benefiting by buying (importing) commodities at low prices from peripheral exporting countries.
Altogether, their peripherality in the WTN and their lack of import-brokerage power may explain why these countries do not meet the economists' expectations (see \url{http://www.businessinsider.com/why-china-was-the-only-bric-to-succeed-2013-10?IR=T}).

\subsection*{Biology: metabolic networks}

\paragraph{Linking wiring to function in metabolic networks.}
We use our directed metabolic networks and investigate if enzymes involved in similar node roles (i.e., similar chains of metabolic reactions) perform similar biological functions. We use Gene Ontology (GO) \cite{ashburner00} to annotate enzyme-coding genes with their biological functions. We downloaded the gene-to-GO term mappings from NCBI (\url{ftp://ftp.ncbi.nlm.nih.gov/gene/DATA/gene2go.gz}) in March 2015 and only use the experimentally confirmed GO annotations. Each GO term represents either a biological process (BP), a molecular function (MF) or a cellular component (CC).

We cluster nodes/enzymes of the human metabolic network using Chavl {\cite{lerman91}}, a publicly available hierarchical ascendant classifier, based on their directed graphlet degree vector similarity. Unlike other common clustering methods, Chavl also proposes cuts of the classification tree  based on likelihood linkage analysis {\cite{lerman91}}.
We use the GO term annotation of the enzymes as a benchmark of their biological functions, cutting GO tree at level 5 to standardise GO annotation  {\cite{singh08,liao09,alkan14}}. 781 enzyme coding genes from human metabolic network have at least one GO annotation at level 5.
For the clusterings proposed by Chavl, we calculate GO term enrichment of the grouped genes using standard model of sampling without replacement {\cite{kuchaiev10}}. We calculate the p-value, corresponding to the probability of obtaining the same or higher enrichment  by chance, using the cumulative hyper-geometric function:
\begin{equation}
p=1-\sum_{i=0} ^{X-1}\frac{\binom{K}{i} \binom{M-K}{N-i}}{\binom{M}{N}},\label{eq:04}
\end{equation}
where $N$ is the size of the cluster (only annotated genes from the cluster are taken into account),  $X$ is the number of genes in the cluster that are annotated with the GO term in question, $M$ is the number of all genes in the network that are annotated with any GO term, and $K$ is the number of genes in the network that are annotated with the  GO term in question.
A cluster is significantly enriched in a given GO term if the corresponding enrichment p-value is smaller than or equal to 0.01.
As presented in Fig. \ref{fig:metabolic} (panel a), we obtain two clusterings: one having 4 clusters, and the other having 19 clusters. All these clusters are significantly enriched in GO terms, showing that enzymes involved in similar node roles in metabolic networks perform similar biological functions.
This suggest that, similar to proteins in undirected PPI networks {\cite{milenkovic08}}, the local wiring patterns around the enzymes in a metabolic network can be used to predict their biological functions, by transferring GO annotations among enzymes with similar directed graphlet degree vectors. 


We use our CCA framework to uncover biological functions that are topologically orthologous (i.e., performed by enzymes involved in similar node roles) between pairs of metabolic networks, of human and four commonly used model organisms: baker's yeast ({\em Saccharomyces cerevisiae}), fruit fly ({\em Drosophila melanogaster}), thale cress ({\em Arabidopsis thaliana}) and mouse ({\em Mus musculus}).
For each species, we create its node role matrix, $R^{t}$, and its GO annotation matrix, $R^{f}$, in the following way.
For both matrices, the same row index represents the same enzyme-coding gene.
In the node role matrix, $R^{t}$, values in a variable vector are the 129 directed graphlet degrees of the enzyme in the corresponding metabolic network, while in the GO annotation matrix, $R^{f}$, they are the Gene Ontology annotations of the corresponding gene (each column correspond to a specific GO term, and the entry is set to 1 if the gene is annotated with the term and 0 otherwise). We considered separately the GO terms describing biological processes, BP, molecular functions, MF, and cellular components, CC, so each species has three different GO annotation matrices. We uncover topologically orthologous biological functions in two steps.
In the first step, we apply our CCA framework between the node role matrix and biological annotation matrix of each species separately and use the corresponding association matrix (which predict the biological annotations of a gene from the directed graphlet degrees of the corresponding enzyme in the metabolic network) to uncover biological functions that are statistically significantly predicted from topology (statistical significance is computed according to 1,000 randomized experiments in which rows in the node role matrix are randomly shuffled; a biological function is significantly predicted if it's p-value, after Benjamini Hochberg correction, is lower than 0.05).
In the second step, biological functions that are statistically significantly predicted in the two species and that have predictions based on similar combinations of directed graphlet degrees, are considered as topologically orthologous.

As presented in Fig. \ref{fig:metabolic} (panel b), human and mouse have the largest number of biological functions that are topologically orthologous. This is expected since human and mouse are the closest species among the five considered ones.
A further validation is that we show that DNA metabolic process, which is one of the essential process for all living organisms, is topologically orthologous across all the considered species (appearing as topologically orthologous in eight out of the ten pairs of species); its significantly associated node roles (orbits) are presented in Fig. \ref{fig:metabolic} (panel c). Our topologically orthologous biological functions may be used to refine the inter-species based function predictions that is currently based only on sequence similarities.
However, a deeper investigation on the meaning of the uncovered relationships between node roles and biological functions is a subject of future research.

\subsection*{Other domains}
Directed network data are abundant in many other domains, e.g. brain networks, e-mail communication networks, citation networks, world wide web networks, and internet peer-to-peer networks, to name a few.  Research on uncovering their organizational principles and function will continue.  Applying our new directed graphlet-based methods, or devising new methods for their analyses {\cite{trpevski2016}} are a subject of future research.

\section*{Concluding remarks}
By generalising the graphlet-based node role framework to directed graphlets, we have enabled advanced descriptive and predictive analyses of directed networked data that could not have been achieved by using undirected graphlets or other directed network statistics.
We have shown that correlations between directed graphlet based node roles, which we capture in the directed graphlet correlation matrices, can be used to sensitively and robustly group networks from the same model and separate those from different models.
Analysing WTNs over years, our framework allows us to: (1) uncover the directed core-broker-periphery organisation of the world trade, in which peripheral countries are specialized in either import or export roles; (2) predict the economic attributes of countries from their positioning in the WTN better than by using traditional directed descriptors or undirected graphlets that may inform regulators about benefits of trade agreements and predict success of an emerging economy; (3) identify relationships between the roles of countries in the WTN and their economic attributes; (4) finely track the dynamics of the WTN,
yielding insights into dominating in trade as brokers of import or export.
Our methodology is general and works in other domains.
We observe that enzymes involved in similar patterns of metabolic reactions in the metabolic network of human are also involved in similar biological functions, and we identify conserved biological functions performed by enzymes that are similarly wired in metabolic reactions across species.
Hence, our directed graphlet based node role framework is universal and promises to deliver insights in the other domains of data science.

\section*{Acknowledgements}
\paragraph{\bf Funding:} This work was supported by the European Research Council
(ERC) Starting Independent Researcher Grant 278212, the
National Science Foundation (NSF) Cyber-Enabled Discovery
and Innovation (CDI) OIA-1028394, the ARRS project J1-5454, and the Serbian Ministry of
Education and Science Project III44006.

\section*{Author contributions statement}

A. S. generalized graphlets, network distances and network models to the directed case and performed the validation on model networks and the study of metabolic networks.
N. M-D. performed the validation against undirected distances and the study of world trade data, and wrote the manuscript.
O. N. Y. helped conducting computations.
N. P. conceived and directed the study.
All authors analysed the results and reviewed the manuscript.

\section*{Additional information}
\textbf{Competing financial interests:} none declared. 

%
%


\newpage

%
%
\begin{figure}
	\begin{center}
	\includegraphics[width=18.3cm]{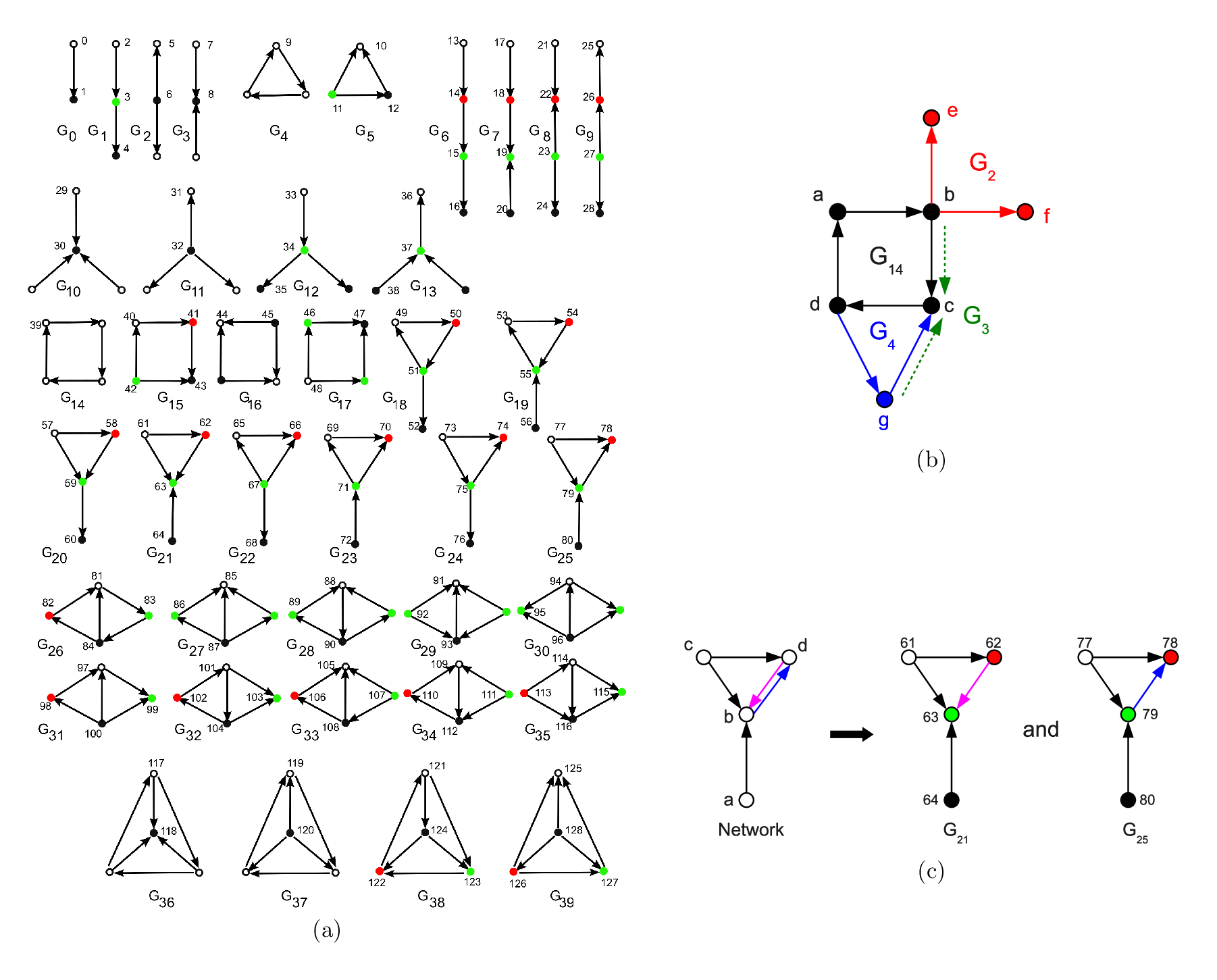}
	\caption{{\bf Illustration of directed graphlets.} (a) The 40 two- to four-node directed graphlets $G_0$, $\dots$, $G_{39}$. In each graphlet, node belonging to the same automorphism orbit are of the same colour.
	The 128 automorphism orbits are labelled from 0 to 127.
	(b) Illustration of how directed graphlets assemble together to form complex networks. The whole network can be created in three steps. First, we start with one graphlet $G_{14}$ (nodes a,b,c and d, in black). Then, we add a graphlet $G_2$ (in red) by adding two new nodes, $e$ and $f$, as heads of directed edges from node $b$. Finally, we add a graphlet $G_4$ (in blue) by adding a new node, $g$, as the head of a directed edge from node $d$ and as the tail of a directed edge towards node $c$. Note that during this process, many new graphlets are created, e.g., graphlet $G_3$ (in green) between nodes $b$, $c$ and $g$.
	(c) Illustration of our anti-parallel directed edge counting strategy. The two anti-parallel directed edges in the input network (in blue and in magenta) account for one graphlet $G_{21}$ (when considering the magenta directed edge) and one graphlet $G_{25}$ (when considering the blue directed edge), among all other induced graphlets. \label{fig:graphlets}}
	\end{center}
\end{figure}

%
%
\begin{figure}
	\begin{center}
	\includegraphics[width=18.3cm]{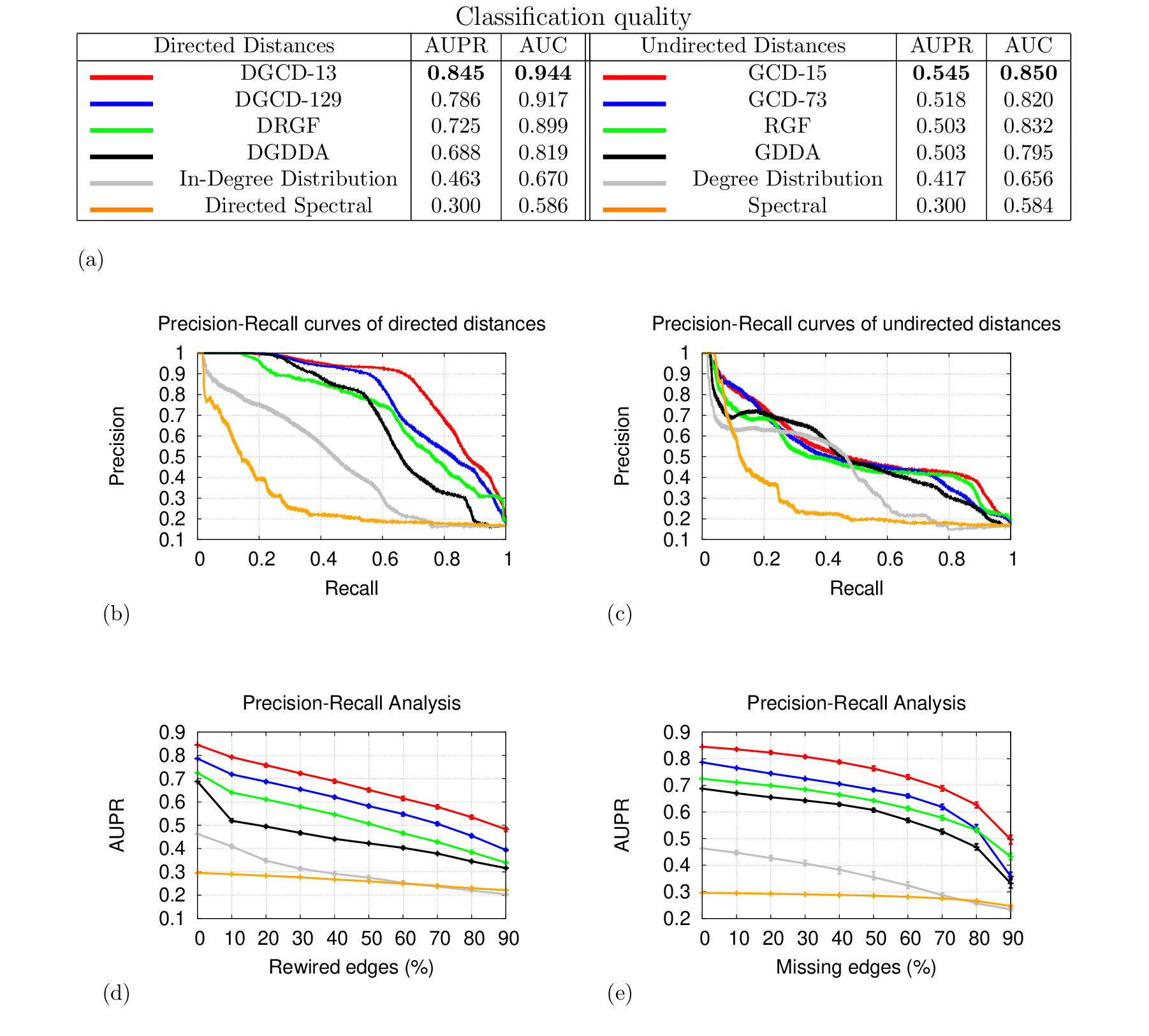}
	\caption{{\bf Quality of clustering of the 360 model networks using directed or undirected distance measures (colour coded in panel a).}
		(a) For directed and undirected distance measures, the area under the ROC curves (AUC) and the area under the precision-recall curve (AUPR) achieved by a distance measure.
		(b) Precision-Recall curves achieved by the six directed distance measures.
		(c) Precision-Recall curves achieved by the six undirected distance measures.
		(d) For directed distance measures, AUPR for different percentages of noise (randomly rewired directed edges in 10\% increments, horizontal axis) in model networks. (e) For directed distance measures, AUPR for different percentages of incompleteness (randomly removed directed edges in 10\% increments, horizontal axis) in model networks.\label{fig:clustering}}
	\end{center}
\end{figure}

%
%
\begin{figure}
	\begin{center}
	\includegraphics[width=18.3cm]{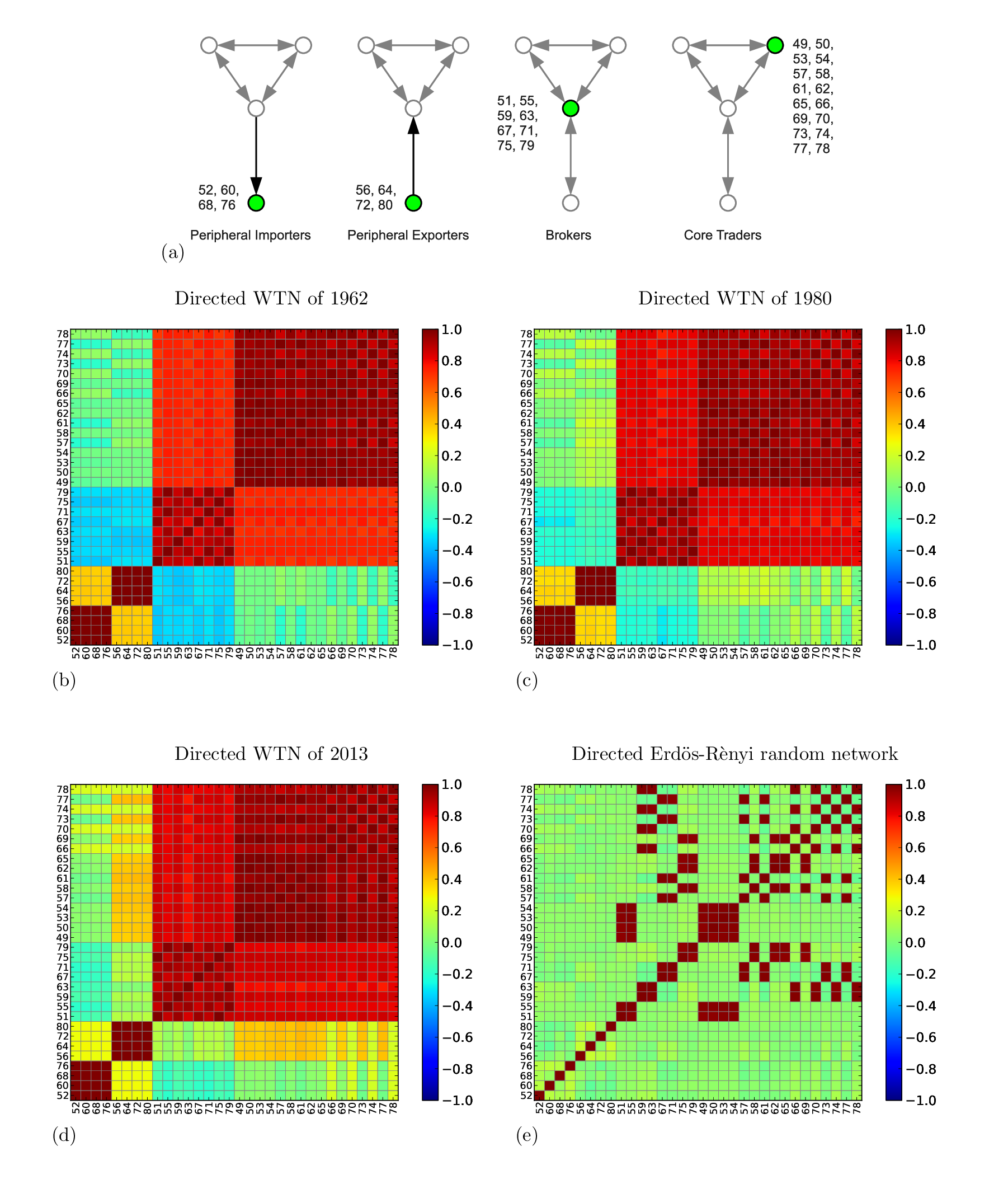}	
	\caption{{\bf The directed Core-Broker-Periphery organization of WTN, with specialized peripheral importers and exporters.}
		(a) The peripheral importer, peripheral exporter, broker and core orbits that we use to build the presented DGCMs. In the graphlets, bi-directional edges (in grey) mean that any directionality is allowed.
		The next panels present the corresponding DGCMs for WTN of 1962 (panel b), WTN of 1980 (panel c), WTN of 2013 (panel d) and for a directed Erd\"os-R\`enyi random network (panel e).
		\label{fig:dwtn_struct}}
	\end{center}
\end{figure}

%
%
\begin{figure}
	\begin{center}
	\includegraphics[width=18cm]{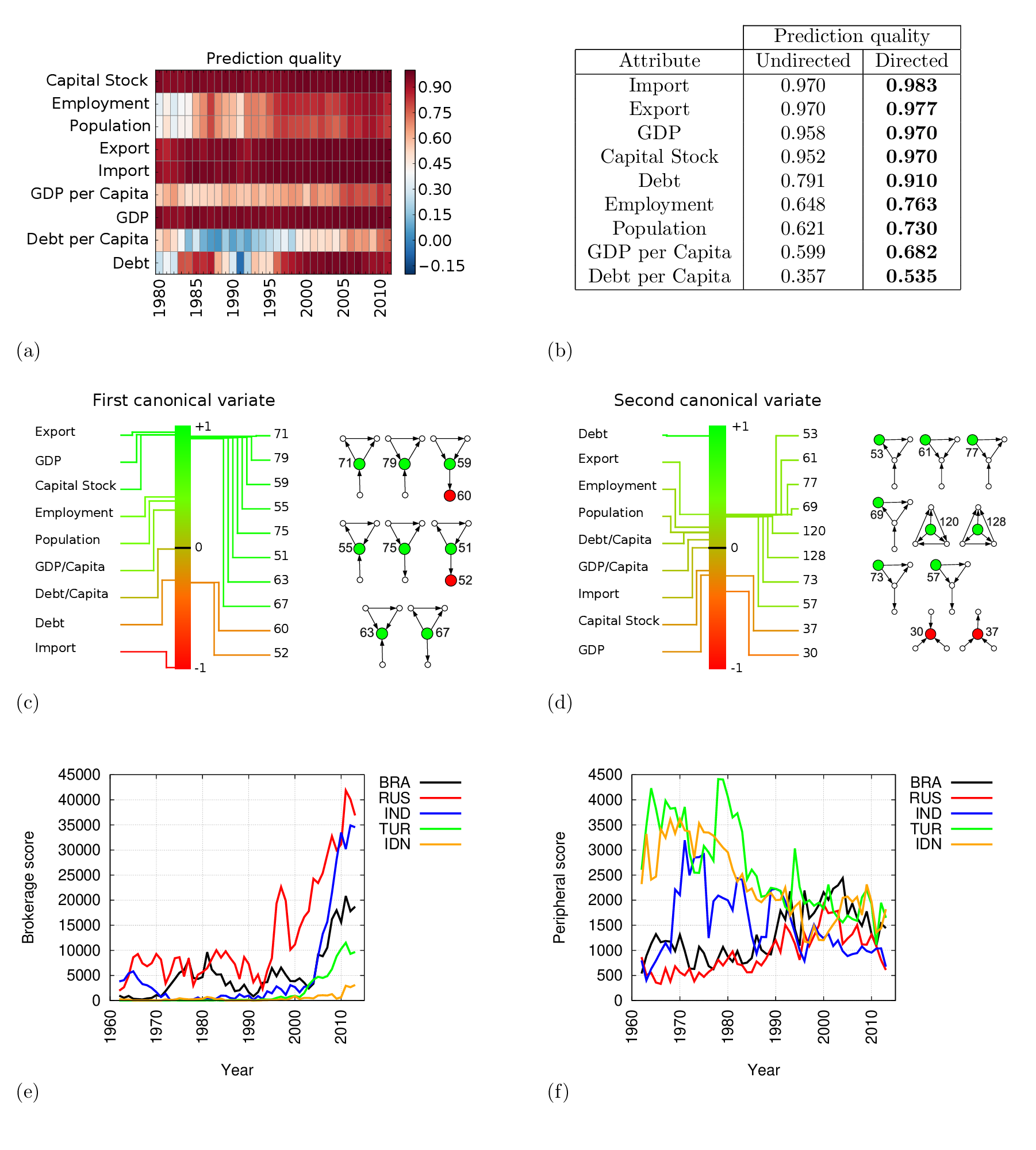}		
	\caption{{\bf Analysis of directed world trade.}
			(a) Yearly assessment of the predictive power of directed graphlets. Economic attributes are predicted from the canonical correlation computed using all countries and all years from 1980 to 2011 simultaneously; the quality of the predictions is measured by Pearson's correlation between the predicted and the real economic attributes of the countries.
			(b) Predictive power of directed- and undirected graphlets. Predictions are computed as in panel a, but the quality of the prediction is measured for each year separately.
			(c) and (d) The first and second canonical variates between economic attributes (on the left) and direct-graphlet orbits (on the right), computed using all countries and all years from 1980 to 2011 simultaneously; the middle bar is colour-coded value of correlation (loading), from -1 (in red) to +1 (in green).
			(e) and (f) Brokerage and peripheral scores of Brazil (BRA), Russia (RUS), India (IND), Turkey (TUR) and Indonesia (IDN).\label{fig:dwtn}}
	\end{center}
\end{figure}

%
%
\begin{figure}
	\begin{center}
	\includegraphics[width=18.3cm]{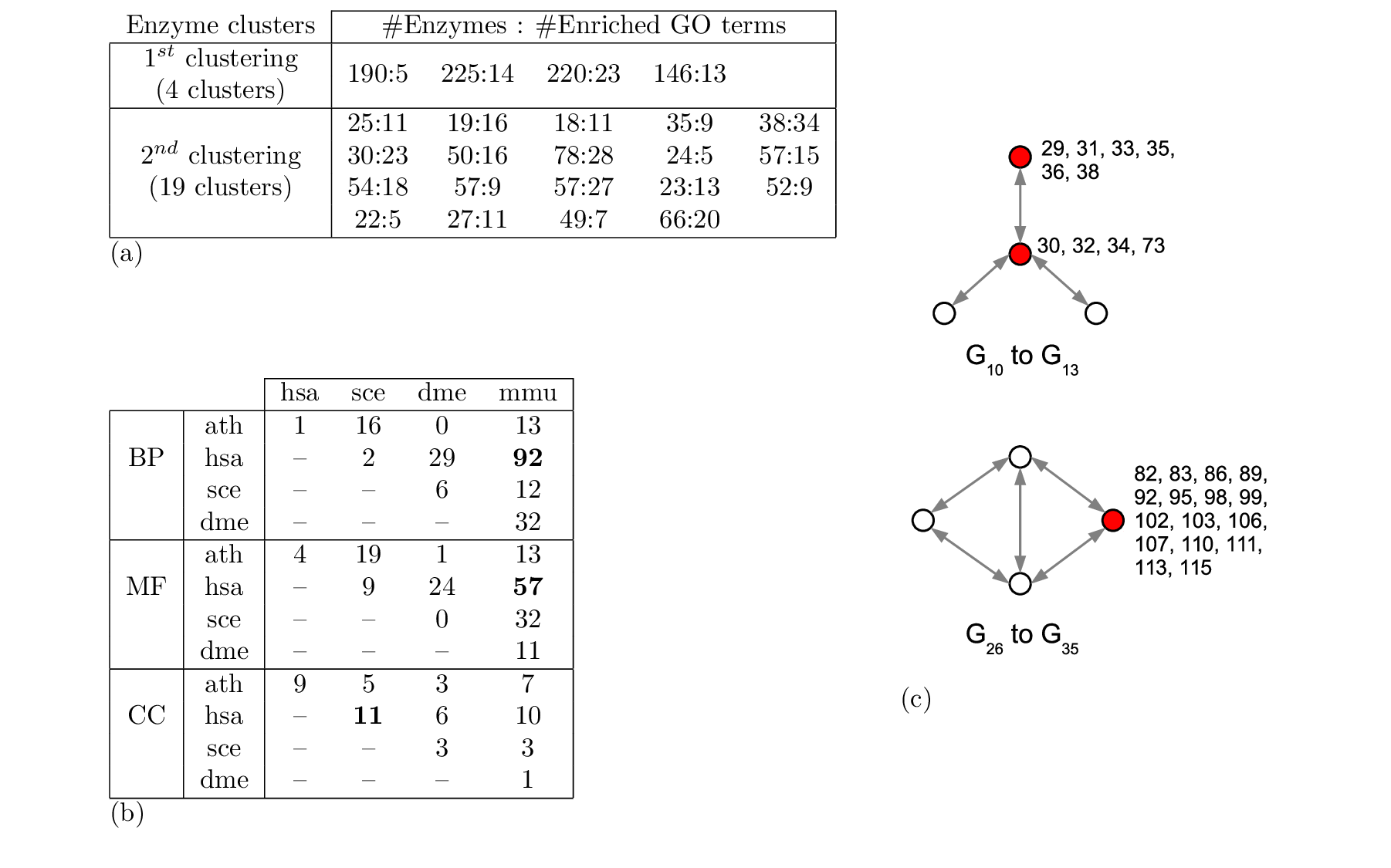}		
	\caption{{\bf Analysis of metabolic networks.}
		(a) GO term enrichment of the two node role based clusters of human enzymes. For each of the two clustering, the GO terms enrichment of each cluster is presented as a pair x:y, where x is the number of enzymes in the cluster and y is the number of enriched GO terms.
		(b) The number of topologically orthologous GO terms per species pairs.
		(c) The significant orbits (in red) for DNA metabolic process annotated enzymes; orbit numbers are presented next to the nodes in red (there are many orbits because our methodology treats different edge directions in separate graphlets).  \label{fig:metabolic}}
	\end{center}
\end{figure}

\end{document}